# Terrestrial Planet Formation During the Migration and Resonance Crossings of the Giant Planets

Patryk Sofia Lykawka[1] and Takashi Ito[2]

[1] Astronomy Group, Faculty of Social and Natural Sciences, Kinki University, Shinkamikosaka 228-3, Higashiosaka-shi, Osaka, 577-0813, Japan; patryksan@gmail.com
[2] National Astronomical Observatory of Japan, Osawa 2-21-1, Mitaka, Tokyo 181-8588, Japan

# ABSTRACT

The newly formed giant planets may have migrated and crossed a number of mutual mean motion resonances (MMRs) when smaller objects (embryos) were accreting to form the terrestrial planets in the planetesimal disk. We investigated the effects of the planetesimal-driven migration of Jupiter and Saturn, and the influence of their mutual 1:2 MMR crossing on terrestrial planet formation for the first time, by performing N-body simulations. These simulations considered distinct timescales of MMR crossing and planet migration. In total, 68 high-resolution simulation runs using 2000 disk planetesimals were performed, which was a significant improvement on previously published results. Even when the effects of the 1:2 MMR crossing and planet migration were included in the system, Venus and Earth analogs (considering both orbits and masses) successfully formed in several runs. In addition, we found that the orbits of planetesimals beyond $a$ ~1.5-2 AU were dynamically depleted by the strengthened sweeping secular resonances associated with Jupiter's and Saturn's more eccentric orbits (relative to present-day) during planet migration.

However, this depletion did not prevent the formation of massive Mars analogs (planets with more than 1.5 times Mars' mass). Although late MMR crossings (at $t > 30$ Myr) could remove such planets, Mars-like small mass planets survived on overly excited orbits (high $e$ and/or $i$), or were completely lost in these systems.

We conclude that the orbital migration and crossing of the mutual 1:2 MMR of Jupiter and Saturn are unlikely to provide suitable orbital conditions for the formation of solar system terrestrial planets. This suggests that to explain Mars' small mass and the absence of other planets between Mars and Jupiter, the outer asteroid belt must have suffered a severe depletion due to interactions with Jupiter/Saturn, or by an alternative mechanism (e.g., rogue super-Earths).

**Key words:** Earth — Methods: numerical — Minor planets, asteroids: general — Planets and satellites: dynamical evolution and stability — Planets and satellites: general — Planets and satellites: individual: Mars

# 1. INTRODUCTION

The presence of significant atmospheres of nebular composition (hydrogen and helium) around all giant planets, and the fact that the nebular gas probably dispersed in less than ~10 Myr (Haisch et al. 2001; Pascucci et al. 2006), strongly suggests that giant planet formation proceeded over a comparable timescale (e.g., a few Myr) to nebula gas dissipation (Lissauer 1987; Pollack et al. 1996; Kenyon 2002; Goldreich et al. 2004). The situation in the inner solar system is less clear, but it is generally believed that the formation of planets in this region occurred on timescales several times longer than for the giant planets. This has been confirmed by a large number of studies on terrestrial planet formation (Lissauer 1993; Chambers & Wetherill 1998; Agnor et al. 1999; Ito & Tanikawa 1999; Chambers 2001; Levison & Agnor 2003; Raymond et al. 2004; O'Brien et al. 2006; Morishima et al. 2008; Morishima et al. 2010; Lunine et al. 2011; Walsh & Morbidelli 2011). Geochemical evidence also suggests that the Earth formed more slowly than Jupiter, in agreement with the above timescale constraints (Kleine et al. 2009).

Models of the early orbital architecture of Jupiter and Saturn suggest that Saturn acquired a semimajor axis ($a$) ~7.2 or ~8.7 AU after being captured in an external mean motion resonance (3:2 or 2:1) with Jupiter, which was supposedly located around 5.5 AU (Morbidelli et al. 2007; Thommes et al. 2008b; Zhang & Zhou 2010). The planets likely migrated to their current locations via interactions with the remnant disk planetesimals at a later time in solar system history (e.g., Fernandez & Ip 1984; Hahn & Malhotra 1999; Levison et al. 2007). The orbital architecture of the asteroid and trans-Neptunian belts adds further evidence for the migration of the giant planets, as mentioned above (Malhotra 1995; Liou & Malhotra 1997; Hahn & Malhotra 2005; Murray-Clay & Chiang 2005; Lykawka & Mukai 2008; Levison et al. 2009; Minton & Malhotra 2009). In this way, Jupiter and Saturn may have experienced migration during or after terrestrial planet formation.

Terrestrial planet formation can be roughly summarized in the following main stages. First, small planetesimals formed with diameters of up to about 100 km (or masses of ~$2.6 \cdot 10^{-7}$ $M_\oplus$) via cumulative accretion or gravitational collapse of smaller bodies (Lissauer 1993; Kokubo & Ida 2000; Chambers 2007; Morbidelli et al. 2009b). Next, a number of planetesimals grew much faster than the background small planetesimals via accretion of the latter, becoming the embryos of the system (the so called 'runaway growth' stage) (Wetherill & Stewart 1989; Ida & Makino 1993; Kokubo & Ida 1998; Rafikov 2003; Goldreich et al. 2004; Leinhardt & Richardson 2005; Morishima et al. 2008; Leinhardt et al. 2009). After this stage, as the population of planetesimals gradually decreases, the growth timescale of embryos increases. The resulting system typically yields tens to hundreds of embryos and a remaining population of small planetesimals. However, when the planetesimals can no longer keep the embryos on dynamically cold orbits (low eccentricity $e$ and inclination $i$), the system becomes unstable. Embryo-embryo collisions and dynamical excitation of planetesimals (i.e., higher $e$ and $i$) then occur during the last stage of terrestrial planet formation (Lissauer 1993; Chambers 2001; Goldreich et al. 2004; Raymond et al. 2004; Chambers 2007; Raymond et al. 2009; Lunine et al. 2011).

Terrestrial planet formation is usually modeled using planetesimal disks set in the inner solar system (<5 AU) consisting solely of embryos, or embryos embedded in a sea of small planetesimals (as discussed in more detail below). Terrestrial planet formation has also been modeled during the first Myr of solar system history. At that point, the nebular disk gas played an important role via gas drag, type-I migration, or gas-driven secular resonance sweeping in the inner solar system (e.g., Kominami & Ida 2004). However, the best examples of these models (Nagasawa et al. 2005; Thommes et al. 2008a) have serious shortcomings, such as too many formed planets (on average, more than the 3-4 planets obtained in other models), atypically long gas-dispersion timescales (timescales longer than 10 Myr are essentially required), or early Moon-forming giant collisions (<20-30 Myr; timescales that are marginally consistent with the currently established constraints at best [see also Section 2]).

Further criticism can also be found in O'Brien et al. (2006), Raymond et al. (2009), Morishima et al. (2010), and Morbidelli et al. (2012). Therefore, we do not discuss these models in this paper.

Additionally, among the scenarios modeled with embryos and planetesimals in a disk, some considered Jupiter/Saturn placed in fixed orbits, and others did not include the giant planets. In summary, all these earlier models generally found that a few Venus- or Earth-like planets were able to form in a region near the orbits occupied by Venus and the Earth, although not on orbits as close as observed. For this reason, one or more such planets often acquired orbits located beyond the Earth and up to $a$ = 2-3 AU. Overall, the results obtained in those studies strongly suggested that planet formation is a highly stochastic process but that, in general, it leads to a relatively stable system of a few final planets after 100-300 Myr (Lissauer 1993; Chambers & Wetherill 1998; Chambers & Wetherill 2001; Raymond et al. 2004; Kokubo et al. 2006; O'Brien et al. 2006; Morishima et al. 2008; Raymond et al. 2009). One typical problem in these models was that the obtained Venus and Earth analog candidates had final eccentricities or inclinations that were too excited (Chambers & Wetherill 1998; Agnor et al. 1999; Chambers 2001; Levison & Agnor 2003; Raymond et al. 2006). One potential solution to this problem is the inclusion of planetesimals in the system, so that dynamical friction could allow the planets to have cold orbits at the end of planet formation. However, none of these models could form Mars analogs *and* solve the problem of water delivery to Earth consistently, among other problems (see Raymond et al. 2004 and O'Brien et al. 2006 for more details).

In an attempt to solve these issues and better explore the initial conditions of Jupiter and Saturn, O'Brien et al. (2006), Raymond et al. (2009), and Morishima et al. (2010) performed detailed simulations where both the giant planets were initially placed on either nearly circular orbits closer to the Sun than now [prior to their migration; the Circular Jupiter–Saturn (CJS) model], or eccentric orbits corresponding to their current orbits [the Eccentric Jupiter–Saturn (EJS) model]. All these studies found that although EJS models could produce systems resembling the solar system terrestrial planets in terms of orbits and masses, the planets obtained acquired hardly any water from the supposedly more water-rich asteroids located beyond 2-2.5 AU. These results confirmed earlier findings discussed in Chambers & Cassen (2002) and Raymond et al. (2004). The CJS models produced planets that were able to incorporate significant amounts of water from outer asteroids, compatible with estimates of Earth's water inventory. Nevertheless, the systems produced in the CJS models typically yielded too massive Mars-like planets, or even planets within the asteroid belt. Nevertheless, these studies found that simulations in which Jupiter and Saturn started on eccentric orbits (comparable to or slightly higher than their current values) tended to produce the best results.

Walsh & Morbidelli (2011) investigated the influence of the migration of Jupiter and Saturn during terrestrial planet formation. The authors considered the setup in which Jupiter and Saturn migrated from 5.4 and 8.7 AU to their current orbits, respectively, using an exponential-folding time of 5 Myr (i.e., $\tau$ = 5 Myr, as detailed in Section 3.2). Their main findings included the formation of Earth- and Venus-like planets, too massive Mars-like planets (thus resembling the main results and implications discussed above), and asteroid belts with *a-i* orbital distributions incompatible with observations.

Other more recent models suggest that the primordial planetesimal disk from which the terrestrial planets formed was much narrower than typically used in several past studies. For instance, Hansen (2009) used disks set at $a$ = 0.7-1.0 AU, obtaining planetary systems similar to the solar system, and Mars-like planets with masses comparable to that of Mars. Inspired by these results, Walsh et al. (2011) proposed that the narrow disk used in Hansen (2009) was created by an early in-then-out gas-driven migration of Jupiter during very early stages of the solar system's history, when the nebular gas was still present. The obtained systems in that model could explain the main orbital and mass distribution of Venus, Earth, and Mars consistently.

The initial orbits of Jupiter and Saturn that were tested in past models, as discussed above, varied from orbits that match the currently known ones (5.20 and 9.56 AU) to those representing their state prior to planet migration[1], where both planets were located closer to the Sun than at present. Internal mean motion resonances (MMRs) with Jupiter and secular resonances associated with the locations of Jupiter and Saturn likely played a significant role in the dynamics of the inner solar system, provided the giant planets had non-zero eccentricities during their migration (e.g., Murray & Dermott 1999). Indeed, the initial eccentricities of Jupiter and Saturn, even within limited ranges (e.g., ~0.01-0.1), strongly affect the main characteristics of the terrestrial planets that form (Chambers & Wetherill 2001; Chambers & Cassen 2002; Raymond et al. 2004; Raymond et al. 2009; Morishima et al. 2010). The timescales for these resonances to sweep the inner solar system also play a significant role in terrestrial planet and asteroid belt formation (Minton & Malhotra 2009; Morbidelli et al. 2010; Minton & Malhotra 2011). In addition, the presence of Jupiter and Saturn can strongly shape the evolution of embryos and planetesimals via gravitational and resonant perturbations, including the inner regions of the planetesimal disk, via energy and angular momentum exchanges among the embryos (i.e., the "secular conduction" discussed in Levison & Agnor 2003).

Given the strong influence of planetary eccentricities and sweeping secular resonances associated with Jupiter and Saturn, we aim to explore the effects of temporarily excited orbits and simultaneous migration for these planets on terrestrial planet formation. The giant planets typically acquire more eccentric orbits in instability models such as the Nice model (e.g., Gomes et al. 2005; Ford & Chiang 2007; Thommes et al. 2008b). Here, we adopt the "standard" Nice model initial conditions as our baseline, to allow straightforward comparison with previous results. We will show that adopting a more compact orbital configuration for Jupiter and Saturn (5.4 and 7.3 AU) leads to essentially identical outcomes for the final system.

There is a long-standing hypothesis that the Moon and terrestrial planets experienced a sudden increase in the impact flux of small bodies several hundred million years after the formation of the planets: the so-called Late Heavy Bombardment (LHB; Hartmann et al. 2000; Strom et al. 2005; Chapman et al. 2007; Bottke et al. 2012). In the Nice model framework, an orbital instability triggered by the mutual 1:2 (or 3:5) MMR crossing of Jupiter and Saturn during their migration is invoked to explain the LHB (Gomes et al. 2005; Morbidelli et al. 2007). However, it is not yet clear if the putative LHB was characterized by an event with intense impact flux in a short timespan (the so-called lunar cataclysm), or a mild and long-standing event as a result of late impactors coming from unstable reservoirs during the first 1 Gyr of solar system history. Although the Nice model predicts a cataclysmic event, detailed analysis of meteorites, lunar rocks, and asteroids suggest that the situation is not yet conclusive. Indeed, even the LHB event itself is under debate (Chapman et al. 2007). Under these circumstances, it is valid to postulate that planet migration and MMR crossings may have happened early in the solar system's history (i.e., within ~100 Myr after the formation of the giant planets), in which case, the LHB would not have been triggered by such mechanisms. Because the terrestrial planets should take around 100 Myr to fully form, it is important to investigate how the dynamical behavior of Jupiter/Saturn under such a scenario may have influenced terrestrial planet formation, and if any new hints can be gained during/after these processes.

Our model is similar to that of Walsh & Morbidelli (2011). However, in addition to the migration of Jupiter and Saturn, we considered that both planets experienced a strong resonance crossing (1:2 MMR here). We also explored the parameter space in much more detail and with a larger number of high-resolution runs (i.e., using 100+ embryos and 2000 planetesimals in each run).

In this way, we modeled the influence of Jupiter and Saturn migration followed by an orbital instability caused by their mutual 1:2 MMR crossing on terrestrial planet formation for the first time. Our simulations add further statistics on terrestrial planet formation, and use a higher number of disk planetesimals than most previous published models. The inner solar system planetesimal disk should be modeled with at least 2000 or 3000 bodies for accurate representation of the system dynamics (Morishima et al. 2008). To date, very few published models used 2000 or more small bodies to represent the primordial inner solar system (Raymond et al. 2006; Morishima et al. 2008; Raymond et al. 2009; Morishima et al. 2010).

---

[1] Planet migration is henceforth referred to as planetesimal-driven migration.

In the scenario envisioned in this work, the giant planets start on nearly circular orbits until the onset of the instability (i.e., the 1:2 MMR crossing event), then later acquire higher eccentricities. In this way, the combined evolution of the low and high eccentricities (and migration) of Jupiter and Saturn possesses the strengths of the CJS and EJS scenarios, as discussed earlier. Furthermore, since MMR crossings involving Jupiter and Saturn can greatly excite the planets' eccentricities, the dynamical perturbations of secular resonances associated with both planets would become stronger. In this way, during the migration of the planets, these strengthened secular resonances could help in clearing the planetesimal disk in the inner solar system as they sweep inwards. In addition, this mechanism could also help to clear the asteroid belt of embryos on quasi-stable (over several hundred Myr) or permanent orbits that would be in conflict with the known orbital architecture of the belt (O'Brien et al. 2007). The depletion caused by Jupiter and Saturn's orbital evolution and migration could also help in explaining the mass discontinuity between the orbits of Earth and Jupiter at the end of planet formation, given the small mass of Mars (e.g., Lissauer 1987).

Concerning the key definitions and notation used in this paper, we define Venus-, Earth-, and Mars-like orbits as follows. Assuming that a planet analog to the Earth would have $a = 1$ AU and averaged $e = 0.03$, and requiring Mars analogs (with averaged $e = 0.07$) not to encounter the Earth by setting the mutual Hill radius > 7 implies a minimum ~1.2 AU for Mars. Therefore, we define Mars-like orbits as those that fall between 1.2 AU and the inner boundary of the asteroid belt at 2 AU. This definition corresponds to the "Mars zone" defined in Chambers & Wetherill (1998). For simplicity, we define Venus/Earth-like orbits as those located within 0.5 and 1.2 AU, requiring both planets to move in stable orbits (i.e., to avoid close encounters). Details on Venus, Earth and Mars analogs are given in Section 2. Finally, unless otherwise stated 'terrestrial planets' will refer to planets obtained in the simulations, and 'planet formation' will refer to terrestrial planet formation.

## 2. MAIN CONSTRAINTS ON TERRESTRIAL PLANET FORMATION IN THE SOLAR SYSTEM

Any successful model of terrestrial planet formation should satisfy a series of constraints. Based on the literature, we have listed typical constraints below as a guide to the level of success of the model. Henceforth, we cite these constraints as "constraint 1", and so on throughout this paper. Although constraints 4, 5, and 6 are not discussed extensively in this paper, we included them for completeness and as reference for future studies.

1) The formation of planets analogous to Mercury, Venus, Earth and Mars within ~ 2 AU (i.e., outside the current asteroid belt) at the end of planet formation. Given the great difficulties in satisfying the constraints associated with each planet, we aim to reproduce the orbits and masses of Venus, Earth and Mars, with an emphasis on Mars. Mercury is not considered in this work, so the terrestrial planet region is defined as $a = 0.5\text{-}2.0$ AU. No published model has fully satisfied this constraint yet, because the formation of Mercury is systematically omitted in previous studies.
Here, for a quick evaluation of the system outcome, we define Venus, Earth and Mars analogs as planets with dynamically cold orbits ($e \sim \sin i < 0.1$) located at 0.5-1.2 AU (Venus-Earth) and 1.2-2.0 AU (Mars) on non-encountering orbits. Their masses must fall within ±50% of currently known values to allow a minimum of 0.05 $M_\oplus$ for an object to be classified a planet (based on Mars' mass), while prohibiting low mass Venus/Earth-like or high-mass Mars-like planets. This translates into 0.4-1.2 $M_\oplus$ for Venus, 0.5-1.5 $M_\oplus$ for Earth, and 0.05-0.16 $M_\oplus$ for Mars. For comparison, Hansen 2009 defines Mars analogs as bodies with 0.02-0.2 $M_\oplus$ located at $a > 1.3$ AU. Here, we consider that if the obtained planets in a certain system *simultaneously* satisfy the range constraints above, the system is deemed acceptable for more detailed analysis using the angular momentum deficit (AMD), radial mass-concentration (RMC), and mutual orbital spacing parameters (see e.g., Chambers 2001).

2) Long-term dynamical stability. Although we assume that terrestrial planet formation ceased after a total time of 200 Myr, we require the planets to survive 1 Gyr for the obtained planetary systems to be considered analogous to the inner solar system. After 1 Gyr, the subsequent orbital evolution is expected to be stable as mechanisms capable of causing major dynamical changes in the orbits of terrestrial planets are not known.

3) Absence of massive bodies in the asteroid belt (defined at $a$ = 2.0-4.5 AU). No planets or massive embryos should remain in the asteroid belt after 1 Gyr, otherwise orbital features not currently seen in the belt arise (e.g., Petit et al. 1999; O'Brien et al. 2007).

4) Origin of the Moon. The giant impact that supposedly created the Earth–Moon system should occur 25-150 Myr after the birth of the solar system (Touboul et al. 2007; Albarede 2009). Similar to Chambers (2007), a giant collision is defined by the collision of an object (impactor) containing at least 10% of the protoplanet's mass (target).

5) Late veneer mass. The amount of material delivered to the Earth via small body impacts after the last giant impact (constraint 4) should not exceed ~1% of the final mass of the Earth (Drake & Righter 2002; O'Brien et al. 2006; Albarede 2009). The biogenic N and C chemical species present on Earth were also likely to have been delivered to our planet after it acquired its bulk mass (Bond et al. 2010).

6) Origin of Earth's water. Sufficient water should be delivered to the Earth from the asteroid belt. This process is very sensitive to the eccentricities of Jupiter and Saturn (Raymond et al. 2004; O'Brien et al. 2006; Raymond et al. 2009). Although an alternative source of water could be short-period comets, current evidence suggests that they were a minor contributor (Hutsemekers et al. 2009). Intriguingly, the delivery of water may have coincided with the accretion of the late veneer mass after $100 \pm 50$ Myr from the birth of the solar system (Albarede 2009).

7) Rapid formation of Mars. Mars likely formed in 1-10 Myr (Nimmo & Kleine 2007). In agreement with that timescale, analysis of Martian meteorites suggest that Mars acquired 90% of its mass in only 2-6 Myr after the birth of the solar system (Dauphas & Pourmand 2011).

We assume Jupiter and Saturn migrated without abrupt orbital changes, such as in the "jumping Jupiter" scenario (Morbidelli et al. 2010), so the associated resonance sweeping of the primordial asteroid belt would represent a secondary constraint, as it likely affected a substantial fraction of the planetesimal disk that provided building blocks during planet formation or, alternatively, provided body impacts on the planets during later stages (e.g., Minton & Malhotra 2011). As Jupiter and Saturn migrated during the early solar system, several internal MMRs associated with Jupiter, the secular resonances $\nu_5$ and $\nu_6$, and other secular and secondary resonances affected the dynamical/collisional evolution of small bodies and embryos in the planetesimal disk at $a$ < 5 AU (e.g., O'Brien et al. 2007). In particular, the $\nu_6$ secular resonance probably played the greatest role in shaping the orbital structure of the planetesimal disk, because it swept the disk from approximately 5 AU to its present location at ~ 2.1 AU (e.g., Chambers & Wetherill 2001; O'Brien et al. 2006; O'Brien et al. 2007; Raymond et al. 2009; But, note that little secular sweeping would occur in "jumping Jupiter" scenarios; e.g., Morbidelli et al. 2010; Nesvorny 2011). The $\nu_6$ secular resonance occurs when the frequency of longitude of perihelion of an object coincides with the fundamental frequency associated with Saturn (Yoshikawa 1987; Morbidelli & Henrard 1991; Murray & Dermott 1999). Objects on initially cold orbits swept by this resonance typically acquire high eccentricities, so that they can encounter Jupiter and other planets, or collide with planets or the Sun. The $\nu_6$ resonance probably swept the primordial asteroid belt in less than 10-20 Myr (O'Brien et al. 2006; Minton & Malhotra 2011).

Other important secondary constraints include the orbital evolution of Jupiter and Saturn. We excluded runs from the analysis if Jupiter and Saturn acquired $e \sim \sin i > 0.15$ to avoid orbital instabilities caused by their small mutual distances. We also excluded runs in which Saturn crossed its mutual 2:5 MMR with Jupiter, since an unknown mechanism would be required to make both planets move back inside that resonance. The current orbital architecture and mass of the asteroid belt provide more constraints, but this is beyond the scope of this paper and will not be discussed here.

In this paper, we emphasize constraints 1-3 and 7. Once these constraints are fully satisfied in upcoming improved models, we intend to explore constraints 4-6 in more detail in future work.

## 3. METHODS

### 3.1 Initial conditions

The starting conditions of our numerical simulations are assumed to represent the solar system when the disk gas became negligibly small, as compared to the mass in solids, corresponding to roughly a few Myr (<10 Myr) after the birth of our system. Overall, the model setup is essentially based on the CJS, EJS, and EEJS ("Extra Eccentric Jupiter/Saturn", where $e = 0.07-0.1$) scenarios, as described in Raymond et al. (2009). In particular, we consider a compact primordial system consisting of Jupiter and Saturn, and a planetesimal disk consisting of hundreds of embryos and thousands of small remnant planetesimals. As commonly adopted in previous studies on terrestrial planet formation, we also assumed that the influences of Uranus and Neptune were negligible, so neither planet was included in the model. Jupiter, Saturn and embryos are considered massive bodies that fully interact with each other, including collisions. Planetesimals are also considered massive and feel the perturbations from the planets and embryos, but they do not interact with each other. Collisions were considered perfectly inelastic, leading to merging of objects when colliding with embryos. Therefore, the final masses of the obtained planets in this work are upper limits, given that in reality accretion efficiency should not be perfect (Genda et al. 2012). However, it is worth noting that fragmentation plays a minor influence on the fundamental properties of obtained planets, such as their number, masses, orbital elements, and growth timescales (Kokubo & Genda 2010).

The standard initial conditions of the systems explored in this work are as follows. We placed Jupiter (J) and Saturn (S) at $a_{J0} = 5.45$ AU and $a_{S0} = 8.18$ AU with $e_{J0} = e_{S0} \sim 2i_{J0} = 2i_{S0} \sim 0.001$. We also tested a more compact system with initial 5.4 and 7.3 AU for both giant planets (akin to the "Jupiter/Saturn in Resonance" (JSRES) scenario in Raymond et al. 2009), but found no significant differences in the outcomes, as discussed in Section 4.2.1. We created a disk of objects representing the planetesimals and embryos formed in the inner solar system. The disk was initially set at 0.5-4.5 AU with a total mass of 5.4 $M_\oplus$ equally distributed between 2000 planetesimals and 100 embryos. Thus, each planetesimal was ~1/10 as massive as the Moon, and the embryos typically started the simulations with masses that were tenths of the mass of Mars. In a special set of runs we increased the number of embryos to 400. The masses of embryos increased with distance from the Sun, as typically obtained in models of runaway growth (e.g., Kokubo & Ida 2000; Leinhardt & Richardson 2005; Kokubo et al. 2006), and used in several past models (Chambers & Wetherill 1998; Raymond et al. 2004; Raymond et al. 2009). Overall, the distribution of mass in the disk resulted in a surface density that obeyed a decay law with exponent -1.5, in agreement with Minimum Mass Solar Nebula (MMSN)-like models of primordial protoplanetary disks (e.g., Hayashi et al. 1985; Lissauer 1987). The distribution of embryos and planetesimals described above is believed to represent the early inner solar system when the effects of the nebular gas were negligible (Agnor et al. 1999; Kokubo & Ida 2000; Chambers 2001; Agnor & Asphaug 2004; Raymond et al. 2004; McNeil et al. 2005; Kenyon &

Bromley 2006; O'Brien et al. 2006; Raymond et al. 2006; Chambers 2007; Morishima et al. 2008; Morishima et al. 2010). The bulk density of planetesimals and embryos was 3 g cm$^{-3}$. Typical initial conditions for the planetesimals and embryos are illustrated in Fig. 1.

Planetesimals started in dynamically cold orbits with $e_0 < 0.01$, $i_0 \sim e/2$, while the embryos started with $e_0 < 0.01$-$0.02$ and $i_0 \sim e/2$. The values of initial eccentricities and inclinations were randomly chosen within these ranges. The initial semimajor axis of the embryos was determined by finding their adjacent distances in terms of their mutual Hill radii, according to $a_{n+1} = a_n + bR_H$. The mutual Hill radius is given by (e.g., Chambers & Wetherill 1998)

$$R_H = \left(\frac{a_{n+1} + a_n}{2}\right)\left(\frac{m_{n+1} + m_n}{3M}\right)^{\frac{1}{3}},$$

where $m$ is the mass of any two adjacent embryos, and $M$ is the mass of the Sun. The parameter $b = 4 + F$, represents a factor in the spacing of the embryos, where $F$ was chosen randomly between 3.5 and 4, so $b = 7.5$-$8$. These values for $b$ were compatible with those found from embryo formation studies and were commonly used in previous terrestrial planet formation models (e.g., Kokubo et al. 2006; O'Brien et al. 2006; Raymond et al. 2009). For the 400 embryos case, $b = 1.7 + G$, where G was generated randomly from 0 to 1.3, so $b = 1.7$-$3$. These values were smaller than the "standard" ones described above. However, this procedure was unavoidable because we required 400 bodies to fit within the modeled 0.5-4.5 AU range. Finally, the three angular orbital elements were chosen randomly in the 0 to 360 degree range for all planetesimals and embryos. All calculations were performed using the MERCURY integrator (Chambers 1999), which has been shown to provide robust and accurate results for planet formation after detailed tests (Raymond et al. 2011).

### 3.2 Planet migration and system orbital evolution

The orbital evolution of the systems described in the previous section was followed according to two main phases. In the first phase, we employed code techniques to mimic the migration of Jupiter and Saturn and their mutual 1:2 MMR crossings. The second phase comprised the long-term evolution of the system after the end of planet migration. We justify this model approach as follows. High-resolution simulations of terrestrial planet formation (as performed here) are computationally very expensive, so it was not feasible to include outer solar system planetesimals (located beyond 4.5 AU) that drive the migration of the giant planets. However, the forced migration of Jupiter and Saturn described below is an acceptable approach (e.g., Malhotra 1995; Hahn & Malhotra 2005; Murray-Clay & Chiang 2005; Lykawka & Mukai 2008; Brasser et al. 2009; Levison et al. 2009; Minton & Malhotra 2009; Lykawka & Horner 2010; Walsh & Morbidelli 2011). In addition, the dynamical outcomes resulting from MMR crossings are strongly chaotic and unpredictable, which can lead to various outcomes, such as too high (e.g., $e > 0.15$) or too small orbital excitation, irregular migration (e.g., with Saturn ending beyond its current orbit), or unstable systems in general (e.g., Tsiganis et al. 2005; Nesvorny et al. 2007). Modeling this would require at least hundreds of terrestrial planet formation runs to provide reliable statistics, which is unreasonable given current computational resources. We overcame this problem by employing the approach described below, which allowed us to model the 1:2 MMR crossing in a more tractable way. Note that for the same reasons, a similar technique was adopted in Brasser et al. (2009) and Morbidelli et al. (2009a).

First, Jupiter and Saturn were forced to move at a constant rate until the orbital period ratio of the planets $P_S/P_J$ was ~1.95. This set the moment preceding their 1:2 MMR crossing. In other words, Jupiter and Saturn slowly approached orbits at ~5.4 and ~8.4 AU, respectively. We tested five distinct timescales for the timing of the 1:2 MMR crossing, namely $t_{RC} = 1, 5, 10, 30$ or 50 Myr, to cover the key times during the formation of the terrestrial planets. Early crossings (1-10 Myr) are expected to deplete the outer regions of the planetesimal disk during the formation of Mars, thus presumably helping to create small-mass Mars-like planets, whereas late crossings (30 and 50 Myr) could yield similar planets as surviving embryos in the system after the depletion of more massive planets formed beyond 1.5 AU.

Next, we took the orbital states of all bodies in the system when $P_S/P_J$ reached ~1.95 as the initial conditions for the planet migration phase. Jupiter and Saturn were forced to migrate exponentially until they acquired their current orbits at 5.2 and 9.56 AU ($P_S/P_J \sim 2.49$). We used a modified version of the MERCURY integrator (Chambers 1999; Hahn & Malhotra 2005) that included planet migration and orbital decay routines (implemented by J. Hahn) for all simulations performed in this work. In the calculations, the semimajor axis of Jupiter and Saturn evolved following

$$a_k(t) = a_{k,final} - \Delta a_k \exp(-t/\tau),$$

where $a_k(t)$ is the semimajor axis of the planet after time $t$, $a_{k,final}$ is the final value of the semimajor axis, $\Delta a_k$ is the displacement for the planet to reach its final orbit, $\tau$ is a constant determining the rate of planet migration, and the index $k$ refers to Jupiter or Saturn. More details on this technique and code implementation can be found in Hahn & Malhotra (2005). The planets migrated over a timescale $t_{mig} = 5\tau$. We tested $\tau = 1$, 2 or 5 Myr, yielding $t_{mig} = 5$, 10 or 25 Myr as the total timescale of the migration phase described above.

During planet migration, Jupiter and Saturn naturally crossed their 1:2 MMR. As discussed earlier, given the chaotic behavior of resonance crossings, we applied a tweak in the model that allowed us to perform a systematic investigation of "successful" outcomes of the 1:2 MMR crossing. In particular, we used an orbital decay routine to damp the eccentricities of both planets such that $(de/dt)_k \propto -\beta e_k$ (where $\beta$ is a constant), and allowed eccentricity and inclination excitation for both planets at the beginning of their migration (typically $e_0 \sim 0.04$-$0.06$ and $i_0 \sim 1$-$1.5$ deg). The eccentricity damping is an expected result of the interactions of outer solar system planetesimals with the planets (e.g., Chambers & Wetherill 2001; Chambers & Cassen 2002; Levison et al. 2007; Morbidelli et al. 2009a). In most systems, Jupiter and Saturn acquired averaged eccentricities of ~0.05 and ~0.10 over a short time span (during the first ~1$\tau$ of migration). We also included systems in which Saturn acquired a more eccentric orbit temporarily ($<e_S> \sim 0.1$-$0.15$), because the associated sweeping ν6 resonance became stronger. A strengthened ν6 resonance could deplete materials in the region near Mars more efficiently, which could explain the planet's low mass (e.g., Morbidelli & Henrard 1991; O'Brien et al. 2006). Another reason for this choice was that, as discussed earlier, more eccentric Jupiter and Saturn orbits were favored in the most detailed past published models, particularly the EEJS scenario. A representative example of the orbital evolution of Jupiter and Saturn as modeled in this work is shown in Fig. 2.

Last, after Jupiter and Saturn acquired their final orbits at the end of planet migration, we let the entire system with all planets and remaining embryos/planetesimals evolve until 1 Gyr. No migration or eccentricity damping was imposed on any particular body during this last phase. The embryos and planetesimals were consistently followed during all phases of the simulations.

In total, 56 runs were performed to cover the key parameters described above, comprising at least three runs of each given scenario (We tested 14 scenarios, represented by Sim1–14 in Table 1). To better understand the role of Jupiter/Saturn migration and their 1:2 MMR effects on terrestrial planet formation, and to facilitate comparison of key results with previous published models, we also performed 12 extra runs in which Jupiter and Saturn evolved on fixed pre-migration orbits over 1 Gyr of the integrations. The basic time step used was 1/60 yr (~6 days). Bodies that acquired heliocentric distances smaller than 0.1 AU and larger than 20 AU were eliminated from the simulations. Particles achieved heliocentric distances greater than 20 AU through gravitational scattering by Jupiter. If these particles were allowed to evolve beyond 20 AU, the great majority would be rapidly ejected from the solar system by further interactions with Jupiter. Furthermore, as discussed later, the formation of terrestrial planets proceeded by accretion of embryos and planetesimals evolving in close orbits at $a < 5$-$10$ AU. Therefore, the influence of the cutoff at 20 AU should be negligible for the results obtained in our simulations. Finally, Table 1 gives a summary of the main parameters of the simulations performed in this work.

## 4. MAIN RESULTS AND DISCUSSION

In this section we discuss the overall results obtained from our 68 high-resolution simulation runs that covered key aspects of terrestrial planet formation in the early solar system. Sim0 covered the non-migrating scenario of terrestrial planet formation, and Sim1–14 explored the migrating systems described in Section 3.2 (Table 1). All system outcomes discussed refer to the state of the system after evolving a total of 200 Myr, after which time we assumed that planet formation had ceased.

### 4.1 Terrestrial planet formation without planet migration

Given that non-migrating scenarios of terrestrial planet formation have been modeled by several groups, we briefly discuss our main results, referred to as Sim0, below. For reference, Sim0 is a typical representative of CJS-type scenarios.

Figure 3 illustrates the time evolution of a typical system modeled in Sim0, which will serve as a guide in the following discussion. In general, the embedded embryos stir each other and slowly grow by accreting planetesimals or other embryos. In the first 10 Myr of evolution (panels a-c), several embryos evolved on more excited orbits, while the planetesimals acquired quite high eccentricities due to perturbations from the embryos and the giant planets (Jupiter and Saturn). The location of the $\nu 6$ resonance was approximately at 4.5 AU and remained fixed over the entire system evolution, because Jupiter and Saturn did not migrate. During the next tens of Myr of evolution (panels d-e), a number of embryos grew larger to form a few planets in the inner regions of the planetesimal disk at $a < 2$ AU. These newly formed planets kept relatively stable orbits due to dynamical friction, which kept their eccentricities small. Because the number (total mass) of planetesimals gradually decreased as the system evolved, the efficiency of dynamical friction also decreased with time. This caused embryos and small mass planets to stir each other and evolve to crossing orbits; so, after tens of Myr of evolution, the system experienced several planet-planet or planet-embryo giant collisions. In the solar system, one of such giant collisions is thought to have occurred to create the Moon (constraint 4 in Section 2. See also, e.g., Ida et al. 1997; Agnor et al. 1999; Canup 2004; O'Brien et al. 2006; Canup 2008).

Beyond 100 Myr (panels f-h), the obtained planets evolved on stable orbits until 1 Gyr due to their sufficiently large mutual distances. However, in a few systems, planets acquired unstable orbits and were removed by Jupiter's gravitational scattering or collided with a planet. Specifically, we found that, in total, eight planets were lost in four runs with timescales peaking around 200–230 Myr, and only two of these planets were lost exceptionally late (after 665 and 685 Myr) in their systems.

Figure 4 summarizes the results of the 12 runs of Sim0 combined, where the obtained planets after 200 Myr are compared to the solar system planets. Overall, Venus- and Earth-like planets appear to concentrate in the region where Venus and Earth are found, showing similar masses and cold orbits. However, this trend clearly extends in the semimajor axis from 0.5 to 2 AU, a feature that conflicts with the observed orbital spacing and mass distribution of Venus, Earth, and Mars. In general, two to five high-mass planets similar to Venus or the Earth can form in a system, but they clearly spread over a region much wider than that observed in the solar system. Besides, on average, three such high-mass planets were obtained, suggesting an extra undesired Venus/Earth-like planet.

Furthermore, all obtained planets located on Mars-like orbits at $a \sim 1.2$-2 AU were 5-10 times as massive as Mars, so latter analogs were not obtained in Sim0. These results confirmed the well-known difficulties faced by non-migrating scenarios in forming Venus and the Earth at 0.7-1 AU (high concentration of mass in a narrow zone) and a small-mass Mars at $a \sim 1.5$ AU (Lissauer 1987; Chambers & Wetherill 1998; Chambers 2001; Kokubo et al. 2006; Morishima et al. 2008; Raymond et al. 2009). Other problems include the survival of small planets and embryos beyond 2 AU, even after 1 Gyr, which is in conflict with observations.

It is important to understand how Mars-like planets dynamically evolve and acquire their final masses, as this can yield important insights into the formation of Mars and evolution of the inner solar system. We identified 11 Mars-like planets from the results of Sim0. As shown in detail in Fig. 6, ten (seven) of these planets originated and acquired most of their masses while wandering in the region around 1-2.5 AU (1.5-2.5 AU). In addition, 9 of the 11 obtained planets started the simulations as seed embryos located within a narrower region between ~1.1 and 2.1 AU, while the remaining two seed embryos were initially located at 0.85 and ~2.9 AU. No correlations of final mass and semimajor axis ranges were seen.

Figure 6 also shows that the majority of the 11 Mars-like planets acquired masses equivalent to that of Mars very quickly, within only a few Myr. Thus, one way to avoid forming Mars analogs that are too massive would be to invoke a mechanism to deplete the source regions of the planetesimal disk at $a$ >1-1.5 AU within 10 Myr from our $t = 0$ (Section 3.1).

The main results, implications and associated difficulties discussed in this section are in excellent agreement with several past studies (e.g., Chambers & Wetherill 1998; O'Brien et al. 2006; Raymond et al. 2009; Morbidelli et al. 2012). Because we performed simulations using a higher number of embryos and planetesimals (Section 3.1) than almost all previous models, our results suggest that higher resolution simulations cannot solve the problems stated earlier. However, because the runs of Sim0 and those published elsewhere used embryos with masses representing large fractions of Mars' mass in the main source region of interest (i.e., a = 1.5-2.5 AU; see also Fig. 1), investigations using hundreds of embryos and several thousand planetesimals will provide a definite answer.

**4.2 Terrestrial planet formation during the migration/1:2 MMR crossing of Jupiter and Saturn**

The results of 14 scenarios (Sim1–14) on the influence of migration of Jupiter and Saturn, and the 1:2 MMR crossing of both planets on terrestrial planet formation are summarized below. In these scenarios, we varied the timescales of two main processes: the timing of the resonance crossing $t_{RC}$ and the planet migration time span $t_{mig}$. In Sim1 we also tested a system in which both processes operated "late" (75 Myr from the start of planet formation; see Table 1). In this particular case, the effective MMR crossing timing was 80 Myr from the start of the simulation, while in other simulations this slow convergence towards the 1:2 MMR crossing began immediately, so it describes the crossing timing directly. Below, we discuss some similarities and differences in evolution in the migrating scenarios vs. the non-migrating scenario described in Section 4.1.

Figure 7 shows the time evolution of a representative system modeled in Sim4, where Jupiter and Saturn took 5 Myr to cross their 1:2 MMR limit and a further 5 Myr to migrate to their current orbits. First, because the system in Sim4 took 5 Myr to suffer a dynamical shake up by the resonance crossing and migration of Jupiter and Saturn, the first 5 Myr of evolution are statistically comparable to those obtained in Sim0 (See panels a-b of Figs. 3 and 7). That is, the embryos gradually grew and perturbed each other, but acquired only mild eccentricities/inclinations due to dynamical friction.

The $\nu_6$ resonance stayed around the region at $a \sim 4.3$-4.5 AU during this first 5 Myr of evolution (see also Fig. 2 of Minton & Malhotra 2009). After the Jupiter/Saturn MMR crossing, the migration phase took only 5 Myr and is illustrated in three different moments in Fig. 7 (panels c-e). The position of the ν6 resonance is very sensitive to the mutual locations of Jupiter and Saturn. In our simulations the ν6 resonance moved from its initial location at 4.5 AU (Fig. 7, panel b) to approximately 2.6 AU in only 1 Myr (Fig. 7, panel c). After a further 2 Myr, because Jupiter and Saturn were close to their final orbits, the ν6 resonance moved close to ~2.1 AU (Fig. 7, panel d), remaining at essentially the same location until the end of planet migration (Fig. 7, panel e). Our system outcomes did not reproduce exactly the orbits of Jupiter and Saturn, so we estimated that the final location of the ν6 resonance was at approximately $a = 2.1$-$2.2$ AU. The exact location of the resonance is not relevant, however. As discussed in Minton & Malhotra (2009; 2011) and Morbidelli et al. (2010), one of the most important aspects related to the ν6 resonance is the timescale for its sweep through the primordial planetesimal disk at 2.1-4.5 AU.

In addition, the strength of the ν6 resonance is also sensitive to the planet's eccentricities. Because the 1:2 MMR crossing excited the eccentricities of Jupiter and, more importantly, Saturn, the perturbation effects of the ν6 resonance were enhanced. Although the ν6 resonance sweeping proceeded in less than 3 Myr in our representative case, and in all other simulations that used $t_{mig} = 5$ Myr (Table 1), its enhanced perturbative effects were important. During the first 10 Myr of planet formation, many embryos were removed from the system because of the ν6 resonance's destabilizing effects and the embryos' mutual perturbations. This was in contrast to the non-migrating scenario, in which embryos remained on relatively stable orbits at $< 3.5$ AU. Those embryos that survived typically remained confined to $a < 2$ AU (i.e., near the final location of the ν6 resonance at $a \sim 2.1$-$2.2$ AU). These features can be seen by comparing the system states at 10 Myr in Figs. 3 and 7.

Internal MMRs with Jupiter (e.g., 1:2, 2:5, and 1:3) played a minor role, as compared to the effects of the sweeping ν6 resonance (and to a lesser extent, the ν5 resonance at $a < \sim 1$-$1.5$ AU). In fact, these MMRs are more important on long time scales, because they are associated with depletion structures in the asteroid belt (e.g., Minton & Malhotra 2009).

The subsequent evolution of our representative system ($t > 10$ Myr) resembled the evolution of the systems in the non-migrating scenario during their middle stages ($t >$ a few tens of Myr). The remaining embryos and planetesimals in the system continued to evolve until a few planets had formed. These planets remained in cold orbits if the number of remaining planetesimals was sufficient to drive dynamical friction (this was particularly seen in cases in which $t_{RC} = 1$, 5, or 10 Myr). After 100 Myr the obtained planets tended to evolve on stable orbits given their mutual distances, in particular at $a < 1.5$-$2$ AU (compare panels f in Figs. 3 and 7). This was similar to the non-migrating scenario. However, we found that, in total, 14 planets were lost in 12 runs of eight migration scenarios after the end of planet formation. These planets were removed from their systems after 200-400 Myr, whereas only one planet was lost after 832 Myr in its system.

The main results discussed above also apply to simulations that used distinct values of $t_{RC}$, (30 or 50 Myr), except that the dynamical events would apply at different periods in the system's evolution. For example, if $t_{RC} = 10$ Myr, the system could be described qualitatively as in panel c in Fig. 3, before suffering the dynamical events discussed above. Likewise, in the case of a "late" occurrence of those processes, with $t_{RC} = 30$ or 50 Myr, the states of the system could be illustrated qualitatively by panels d or e in Fig. 3.

Figure 8 summarizes the results of all runs performed for the 14 scenarios described in this section. As in Fig. 4, the planets obtained after a total of 200 Myr of orbital evolution are compared to the solar system planets. Similarly, regarding the results obtained in the non-migrating scenario (Section 4.1), Venus- and Earth-like planets were obtained with similar orbits and masses. One important difference was that these planets were concentrated in a narrower region at $a = 0.5$-$1.5$ AU. These results suggested that Earth and Venus analogs were successfully obtained in most of the runs of Sim1–14. Runs with $t_{RC}$ = 1-10 Myr tended to result in systems that were more compatible with the inner solar system, while those with $t_{RC}$ = 30 or 50 Myr yielded diverse outcomes, some of which resembled the non-migrating scenario.

In general, the terrestrial planets obtained in Mars-like orbits ($a$ = 1.2-2 AU) in Sim1–14 showed an order of magnitude spectrum of masses, from 0.12 $M_\oplus$ to 1.08 $M_\oplus$. We will discuss the best cases in which a relatively small Mars was obtained, and the dependence on the parameters used in Sim1–14, in Section 4.2.1.

In contrast to the non-migrating scenario, small planets and the remaining embryos located beyond 2 AU tended not to survive the 200 Myr of evolution. As discussed earlier, these objects acquired unstable orbits after suffering perturbations from the sweeping ν6 resonance and from other evolving small planets on eccentric orbits in the same region. These objects tended to be ejected from the solar system by gravitational interactions with Jupiter, or to collide with a giant planet or the Sun. A few small planets are still seen after 200 Myr (Fig. 8), but several of them did not survive the full 1 Gyr evolution.

Although the problem of forming a Mars analog persists, a comparison of Figs. 4 and 8 shows that the distributions of planets obtained in the terrestrial planet region and the effective depletion of unwanted massive bodies beyond 2 AU represent a substantial advance, as compared to the non-migrating scenario (Section 4.1). In addition, a few system outcomes seem to resemble the inner solar system in terms of planet orbits and masses (Section 4.2.1). As in the non-migrating scenario (Sim0), the obtained individual systems varied as a result of the stochasticity of planet formation.

### 4.2.1 Planetary systems analogous to the inner solar system

We discuss now individual planetary system outcomes that yielded the best analogs of Venus, Earth, and Mars, with special emphasis on the formation of Mars analogs (Section 2). We required planetary systems analogous to the inner solar system to satisfy constraints 1 and 2 *in the same system*, and also to have a radial mass concentration (RMC) close to, or greater than half that of, the inner solar system (~45). Another system indicator is the angular momentum deficit (AMD), which equals 0.0018 for the inner solar system (e.g., see Chambers 2001 for details on the RMC and AMD indicators). To obtain hints about the processes that may lead to systems similar to our own, we increased the statistics by relaxing the mass condition of Mars analogs to up to 3 times the mass of Mars (0.32 $M_\oplus$) and by not constraining the range of the resulting system AMDs. The best planetary systems obtained after 1 Gyr correspond to four runs: one each from Sim1 (run1: $m_{Mars}$ = 0.12 $M_\oplus$; RMC = 53.0; AMD = 0.0066), Sim4 (run6: $m_{Mars}$ = 0.22 $M_\oplus$; RMC = 43.5; AMD = 0.0035), Sim9 (run3: $m_{Mars}$ = 0.24 $M_\oplus$; RMC = 50.1; AMD = 0.0015), and Sim10 (run1: $m_{Mars}$ = 0.17 $M_\oplus$; RMC = 44.0; AMD = 0.0030) (See Figs. 9 and 10). In these systems, the number, masses and orbital spacings of Venus, Earth, and Mars analogs seemed to be compatible with those observed in our system. In contrast, the non-migrating scenario did not produce systems that satisfied the above conditions. Particularly, the planets in Mars-like orbits were *at least* 5-6 times more massive than Mars; stable planets with masses up to ~1 $M_\oplus$ survived beyond 2 AU in half of the runs (unwanted extra planets); and the median system RMC was 30. In short, it is clear that migrating scenarios offer a substantial step toward satisfying the constraints of terrestrial planet formation. This strongly suggests that a perturbation was necessary before or during terrestrial planet formation.

Concerning the dynamical and accretional evolution of Mars-like planets, we identified four planets that best approached the conditions of a Mars analog (one from each of the systems shown in Figs. 9 and 10). These planets acquired orbits and masses similar to those of Mars, namely $a$ = 1.57-1.72 AU, $e$ = 0.025-0.16, $i$ = 1.8-9.7 deg, and $m$ = 0.12-0.24 $M_\oplus$. As illustrated in Fig. 11 (Section 4.1, see also Fig. 6), the four planets originated and acquired most of their masses while wandering in the region around 1-2.5 AU. This was similar to the case for the Mars-like planets obtained in the non-migrating scenario. Three of these planets started the simulations as seed embryos within a region at ~0.9-1.8 AU, while a single seed embryo was initially located at ~2.8-2.9 AU. Although no correlations of final mass and semimajor axis were seen, the least massive Mars analog obtained originated in the primordial asteroid belt at $a$ ~ 2.8-2.9 AU.

Figure 11 illustrates how the four obtained Mars analog candidates acquired their final masses. As discussed in Section 4.2, perturbations championed by the ν6 resonance led to a depletion of materials that allowed Mars-like planets to acquire less mass than those obtained in non-migrating scenarios (Fig. 6). However, the four planets acquired masses equivalent to that of Mars in less than 10 Myr and up to twice that mass during the following 10-20 Myr of evolution. Thus, the masses of our Mars "analogs" are still a factor of at least two larger than that of Mars. Similar results were obtained in the EJS and EEJS scenarios of Raymond et al. (2009). These results suggest that the Jupiter/Saturn migration/1:2 MMR crossing mechanism invoked to deplete the planetesimal disk at >1-1.5 AU must operate in the very first few Myr after our $t = 0$ or that a different, more effective mechanism (maybe including migration/MMR crossings) is needed. In addition, the alternative scenario for the elimination of massive Mars-like planets during "late" 1:2 MMR crossings (Sim11–13) did not yield obvious improvements, so the results of that scenario resembled those of the non-migrating scenario.

Of the 14 scenarios tested by Sim1–14, which ones provided the best results in light of the seven constraints posed in Section 2? First, the four "best" Mars-like planets were obtained in runs where $t_{RC}$ = 5 or 10 Myr. Three of these runs employed planet migration on a timescale $t_{mig}$ = 5 Myr. The other obtained systems in Sim2 ($t_{RC}$ = 1 Myr; $t_{mig}$ = 5 Myr), Sim5 ($t_{RC}$ = 5 Myr; $t_{mig}$ = 25 Myr), and Sim6 ($t_{RC}$ = 10 Myr; $t_{mig}$ = 5 Myr) also seemed more in line with those constraints than other simulations (Fig. 8). In conclusion, we believe that migrating scenarios in which $t_{RC}$ = 5 or 10 Myr yielded the most promising results. The situation concerning the migration timescale of Jupiter and Saturn is less clear, because an obvious dependence on the value of $t_{mig}$ used (5 or 25 Myr) was not seen. This suggests that the role of $t_{mig}$ was less important than the parameter $t_{RC}$.

In Sim14 we tested a more compact orbital configuration for Jupiter and Saturn ($a_{J0}$ = 5.4 AU and $a_{S0}$ = 7.3 AU) against the standard initial conditions ($a_{J0}$ = 5.45 AU and $a_{S0}$ = 8.18 AU). In this particular simulation, the giant planets suffered an eccentricity excitation (~0.05) due to the 3:5 MMR crossing (and to a lesser extent, the 4:7 MMR) after 3-5 Myr of evolution. However, the system outcomes obtained in Sim14 were qualitatively indistinguishable from those in Sim6 or Sim9 (in which the other parameters were equal) or other simulations (Fig. 8). Although not exhaustive, this suggests that other MMR crossings (such as the 3:5, 4:7, etc.) play a minor role in terrestrial planet formation. Also, it is unclear whether the planetary systems obtained in past work showed any dependence *solely* on the initial semimajor axes of Jupiter and Saturn (e.g., Raymond et al. 2009). Therefore, these facts suggest that the influence of the initial locations of Jupiter and Saturn is much less important than the giant planets' eccentricities or the stochasticity of planet-formation processes.

In Sim9–10, we increased the number of embryos from 100 to 400 to test the influence of the embryo individual mass on the final mass of the Mars-like planets (Section 3.1). The key parameters used in Sim9 and Sim10 were the same as in Sim6 and Sim8, respectively. Comparing Sim9(10) and Sim6(8) in Fig. 8 suggests that the increase in the number of embryos played a negligible role in the final system state. However, two out of our four Mars analog candidates were obtained in Sim9 and Sim10, so we intend to investigate the influence of the embryo individual masses in future work.

Most of our system outcomes are similar to those obtained in Walsh & Morbidelli (2011) (compare their Fig. 2 with Fig. 8). This shows that the role of $t_{mig}$ was minor and that the perturbations of the sweeping ν6 resonance (even if strengthened by the eccentric orbits of Jupiter and Saturn) were not so decisive in producing systems similar to our own.

In a more recent work, Walsh et al. (2011) obtained a few Mars analogs, but it is difficult to evaluate the level of success of that model against the seven main constraints (Section 2), given the lack of detail concerning their obtained systems and their small-number statistics. It is also unclear how their results address the simultaneous formation of Venus, Earth, and Mars analogs and their long-term stability, and the removal of small planets/embryos in the asteroid belt. Despite these uncertainties, in line with our model, it strengthens the case that the outer regions of the planetesimal disk were perturbed early in the history of the inner solar system.

### 4.2.2 Confronting the results with the main constraints of terrestrial planet formation

We briefly recall the seven main constraints of terrestrial planet formation suggested in this paper: 1. planets analogous to Venus, Earth, and Mars in terms of orbital elements and masses, and located within ~ 2 AU; 2. dynamical stability of the planetary system analogous to ours (at least 1 Gyr); 3. absence of planetary bodies in the asteroid belt at 2-4.5 AU; 4. the timing of the giant impact that created the Earth–Moon system; 5. the amount of material delivered to the Earth via impacts after the formation of the Moon; 6. origin of Earth's water and water delivery processes; 7. the timescale for the formation of Mars.

The dynamical effects associated with the 1:2 MMR crossing and the sweeping of the strengthened ν6 resonance helped deplete the outer parts of the planetesimal disk at $a > 1.5$-2 AU, resulting in the formation of close Venus and Earth analogs and promising small mass Mars-like planets. Another typical outcome was the high removal rate of (unwanted) massive objects beyond 2 AU. Therefore, compared to the non-migrating scenario, a notable advance was achieved in satisfying constraints 1, 2, and 3.

Because our initial conditions applied to a few Myr after the birth of the solar system, constraint 4 would require a Moon-forming late giant collision at $t$ ~ 40-140 Myr. In our model, the preferred scenarios of migration/1:2 MMR crossing favor perturbations experienced by the system on timescales of $t_{RC} + t_{mig}$ ~ 10-35 Myr (i.e., 5+5 or 10+25 Myr, respectively). If giant collisions peaked around the 10-35 Myr period, then our results suggest that the giant collision probably occurred at $t$ = 40 Myr (in agreement with the results of O'Brien et al. 2006). If true, this may make constraint 5 more difficult to satisfy, because more time would be available between the last giant impact (e.g., at $t$ = 40 Myr) and the subsequent impacts of remaining objects in the system until the end of planet formation. The migrating scenarios modeled in this work could also satisfy constraint 6. The EJS models discussed in the introduction produce Earth-like planets that are too dry. In the migrating scenarios, however, during the first phase, when the eccentricities of Jupiter and Saturn were very small (a period limited by $t_{RC}$, Table 1), a number of growing massive bodies in the terrestrial planet region accreted water-rich asteroids from the region beyond 2 AU. Finally, the Mars analog candidates obtained in our simulations achieved masses similar to that of Mars in less than 10 Myr, rapidly enough to satisfy constraint 7. However, at the end of planet formation, these planets acquired too much mass, in conflict with constraint 1.

In summary, given the difficulties of satisfying these constraints consistently, further improvements of this model are warranted. We also intend to investigate constraints 4-6 in future work.

### 5. CONCLUSIONS

We explored the influence of planetesimal-driven migration of Jupiter and Saturn, and their 1:2 MMR crossing on the formation of terrestrial planets during the early solar system. As a first step, we prioritized obtaining Venus, Earth, and Mars analogs with both orbits and masses compatible with those in the solar system. The non-migrating scenario of terrestrial planet formation (in which the giant planets do not migrate) and a further 14 migrating scenarios were explored. In particular, we explored the influence of the timing of the MMR crossing event (1, 5, 10, 30, or 50 Myr after the start of planet formation) and the duration of planet migration (5 or 25 Myr).

We found that the non-migrating scenario yielded system outcomes with properties very similar to those obtained in several past works (Morbidelli et al. 2012 and references therein). We confirmed that the formation of too massive Mars analogs and the survival of planets in the asteroid belt (beyond 2 AU) are the main problems in this scenario. Indeed, there is a 50% chance of having a planet on a stable orbit in the asteroid belt at the end of planet formation. Such planets are often seen in the results of a number of previous studies based on the non-migrating scenario.

We found that the 1:2 MMR crossing excited the nearly circular orbits of Jupiter and Saturn to eccentricities slightly higher than those currently observed, which in turn enhanced the strength of the $\nu_6$ secular resonance and other resonances associated with the giant planets. The $\nu_6$ resonance swept the planetesimal disk during planet migration, thus strongly exciting the orbits of planetesimals/embryos in a broad region of the planetesimal disk at ~1.5-4.5 AU. This resulted in the formation of planetary systems with fewer planets on stable orbits within $a < 2$ AU. These systems included Earth and Venus analogs, concentrated in a region at $a = 0.5$-$1.5$ AU. Mars analogs were also obtained, but they acquired masses 1-10 times the mass of Mars. However, we found that four of 56 final systems that produced Earth and Venus analogs also had Mars-like planets that almost fully satisfied Mars' orbital and mass constraints, displaying $a = 1.57$-$1.72$ AU, $e = 0.025$-$0.16$, $i = 1.8$-$9.7$ deg, and $m = 0.12$-$0.24$ $M_\oplus$. Our results also showed that, in general, ~80-90% of Mars-like planets originated somewhere between ~0.9 and 2.1 AU and acquired most of their masses from a similar region (1-2.5 AU).

Our best terrestrial planet system analogs were obtained when the 1:2 MMR crossing occurred at 5 or 10 Myr (rather than 1, 30, or 50 Myr) after the start of planet formation (our $t = 0$) and Jupiter–Saturn migrated over a total of 5 Myr (rather than 25 Myr). Therefore, our results suggest that these mechanisms, which depleted the planetesimal disk at distances beyond 1.5 AU, probably operated during the first few Myr after $t = 0$. The unsatisfactory results of Sim2–3 ($t_{RC} = 1$ Myr) suggest that the depletion in the disk occurred too early (or was not strong enough), so Mars-like planets managed to acquire final masses ~3-5 times the mass of Mars during the subsequent system evolution. Alternatively, a distinct effective depleting mechanism (including or not migration/MMR crossings) would be needed (e.g., dynamically unstable super-Earths or icy embryos from the outer solar system).

In addition, the Jupiter–Saturn migration/1:2 MMR crossing perturbations caused a substantial depletion of planets, embryos, and planetesimals in the 1.5-4.5 AU region via Jupiter's gravitational scattering, inducing hyperbolic orbits or collisions with other massive bodies. Thus, the absence of planets in the asteroid belt is likely to be the result of the migration of Jupiter and Saturn while they were evolving on relatively eccentric orbits.

Unfortunately our model was unable to satisfy all the main constraints in the inner solar system (Section 2). However, considering the paradigm of the non-migrating scenario, the results of our model represent a substantial advance in our understanding of the key processes that operated during terrestrial planet formation. It also strengthens the link between the orbital evolution of Jupiter and Saturn, and the dynamical/accretional evolution in the inner solar system. In the future, we intend to explore other primordial solar system orbital architectures, the role of higher resolution simulations, the influence of the ice line, the formation of Mercury, and the possibility that Mars represents a leftover embryo scattered to its current orbit (as suggested in e.g., Chambers & Wetherill 1998 and Hansen 2009).

**ACKNOWLEDGEMENTS**

We would like to thank the referee for a number of helpful comments, which allowed us to improve this work. All simulations presented in this work were performed using the general-purpose PC cluster at the Center for Computational Astrophysics (CfCA) in the National Astronomical Observatory of Japan (NAOJ). We are thankful for the generous time allocated to run the simulations. This research was supported by Kinki University Research Grant and partially by the NAOJ Research Collaboration Small Grant. TI acknowledges research funding from the Grant-in-Aid of the Ministry of Education, Culture, Sports, Science & Technology in Japan (21540442/2009-2011) and from the JSPS program for the Asia-Africa academic platform (2009-2011) by the Japan Society for the Promotion of Science.

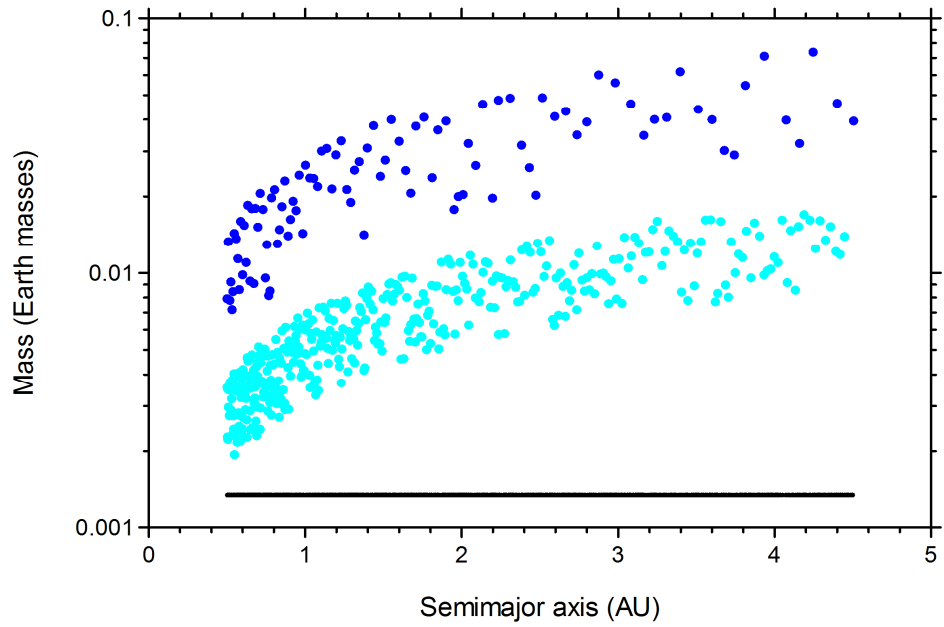

**Figure 1.** Representative initial conditions of embryos and planetesimals used in the simulations. Typically, 100 embryos (blue) and 2000 planetesimals (black, bottom of the figure) were included in the model. In two simulations, 400 embryos (cyan) were employed. See Section 3.1 and Table 1 for more details.

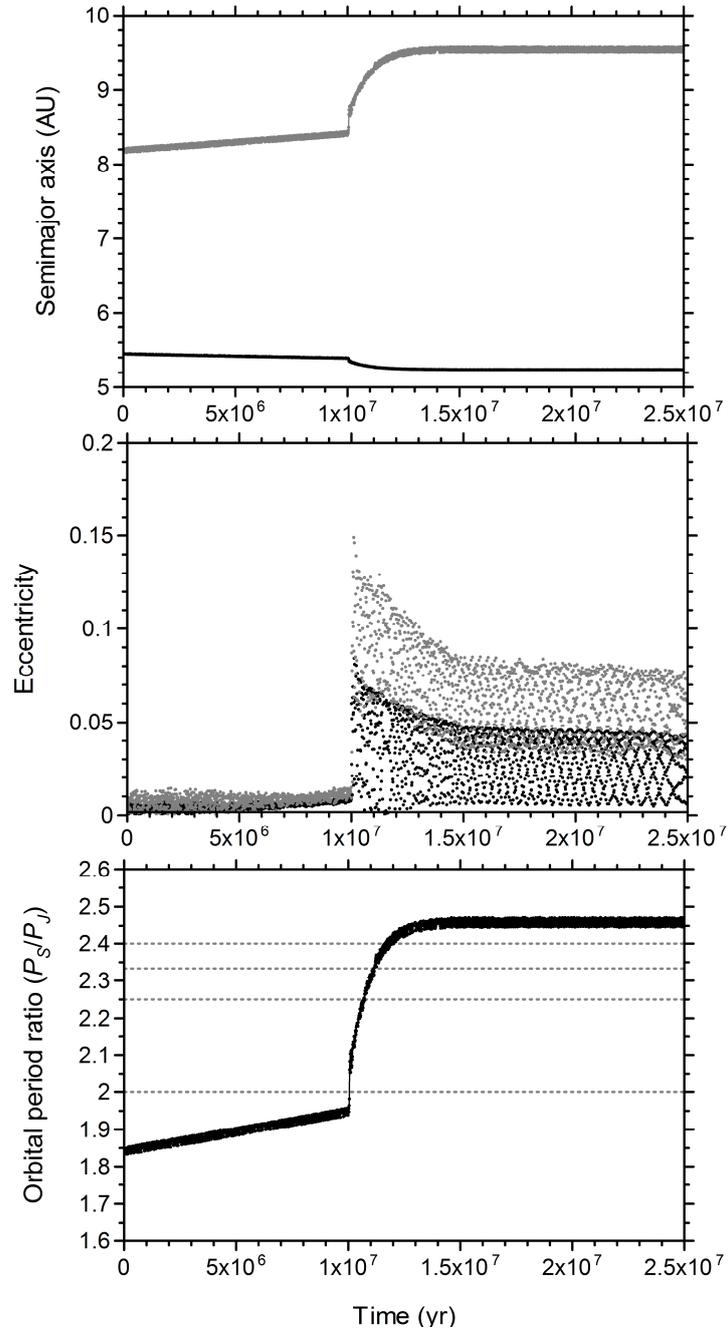

**Figure 2.** Typical orbital evolution of Jupiter and Saturn as modeled in this work (Sim4, run2). In this example, Jupiter and Saturn took 10 Myr to cross their mutual 1:2 MMR, and a further 5 Myr to migrate until both planets reached their final solar system-like orbits. Bottom: Apart from the 1:2 mutual MMR (where $P_S/P_J = 2$), other relevant resonances (3:7, 4:9, and 5:12) are indicated with dashed lines. Similar orbital behavior is seen in other migration models [e.g., Morbidelli et al. (2009a)].

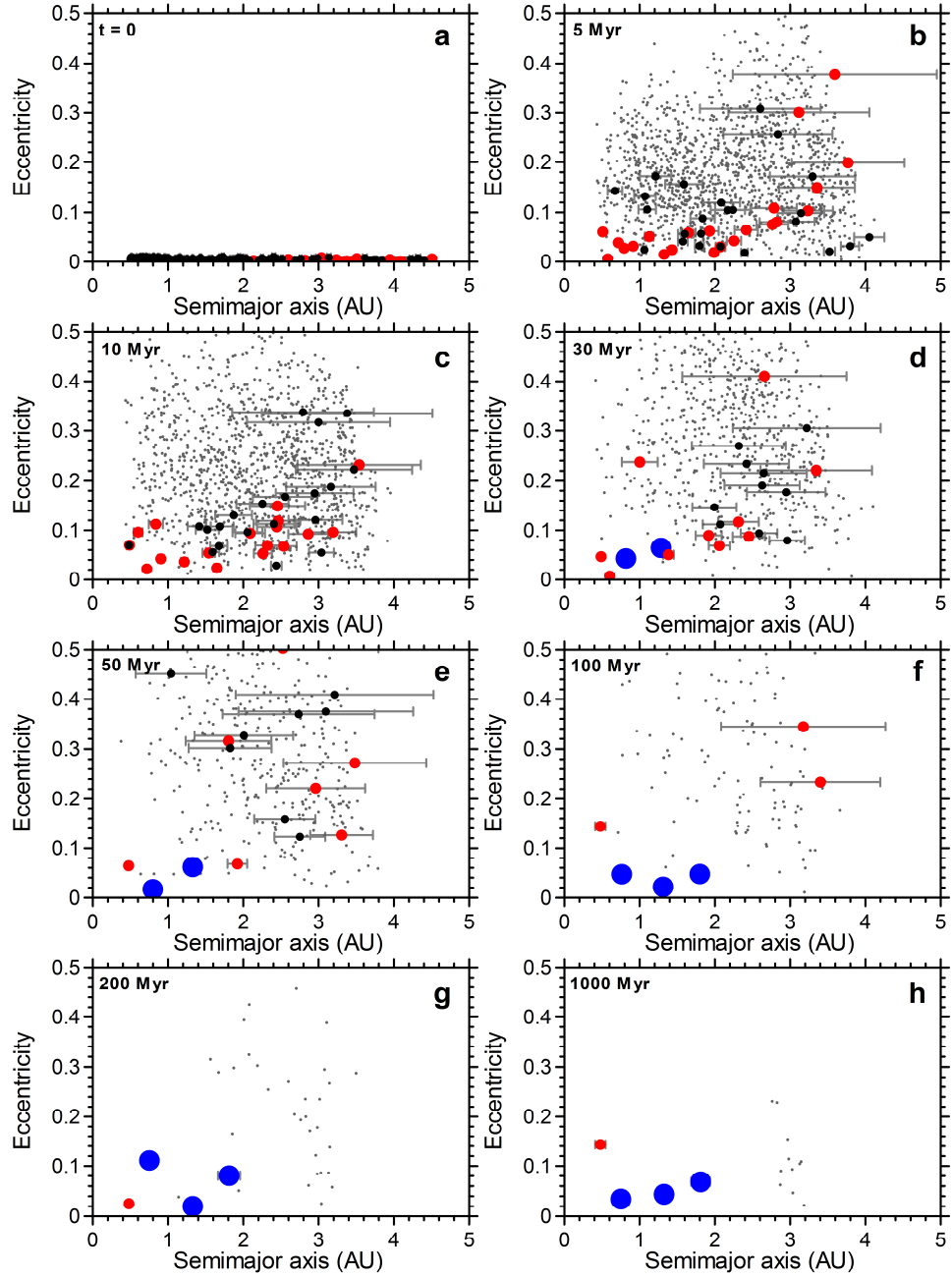

**Figure 3.** Time evolution of a typical non-migrating scenario system, represented by run8 from simulation 0 (Table 1). More massive "Venus/Earth-like" planets (0.5-1.5 $M_\oplus$) are represented by blue circles, while less massive "Mars-like" planets (0.05-0.5 $M_\oplus$) are shown with red circles. Embryos in the system are shown with black circles, while planetesimals are shown with gray dots. Error bars represent the range of heliocentric distances based on the object's perihelion and aphelion. See Section 4.1 for details.

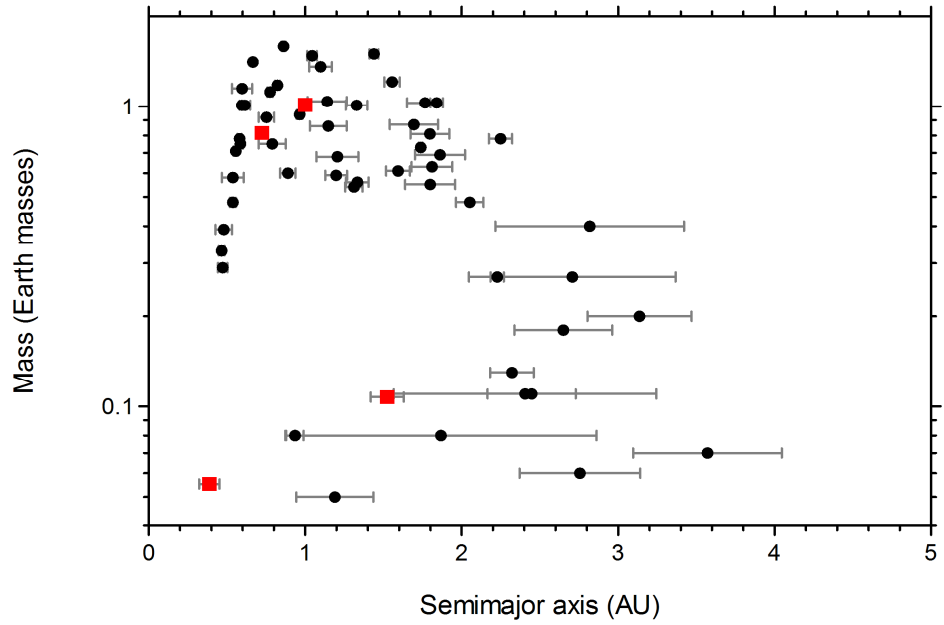

**Figure 4.** Planets obtained in all simulation runs after 200 Myr, representing the general outcomes of the non-migrating scenario "Sim0" (Section 4.1). Red squares represent the four solar system terrestrial planets. Error bars represent the range of heliocentric distances based on the object's perihelion and aphelion.

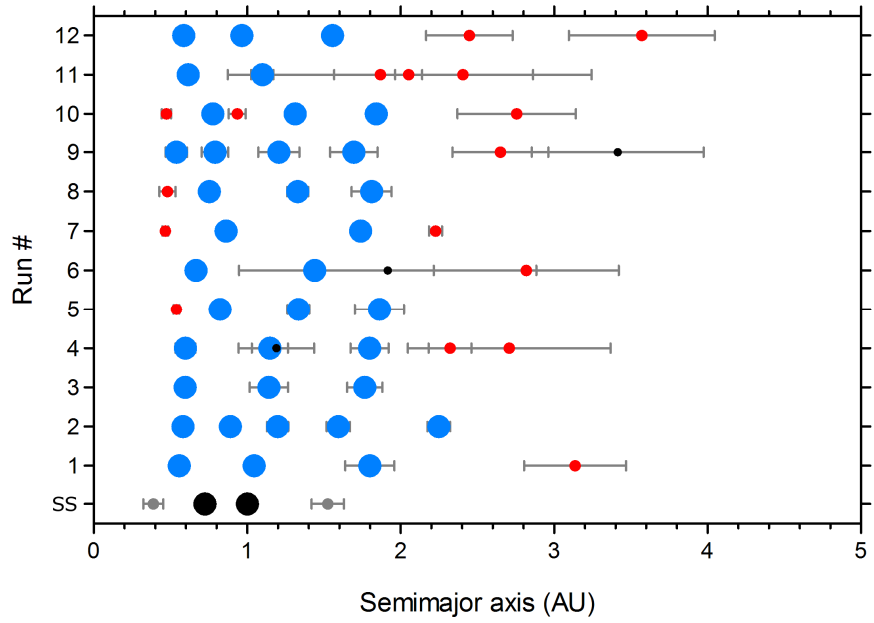

**Figure 5.** Obtained planetary systems after 200 Myr for all simulation runs representing the non-migrating scenario, "Sim0" (Section 4.1). Symbols are the same as explained in the caption of Fig. 3. Error bars represent the range of heliocentric distances based on the object's perihelion and aphelion. (These bars are hidden by the symbols for objects with low eccentricity orbits). The solar system terrestrial planets are shown at the bottom.

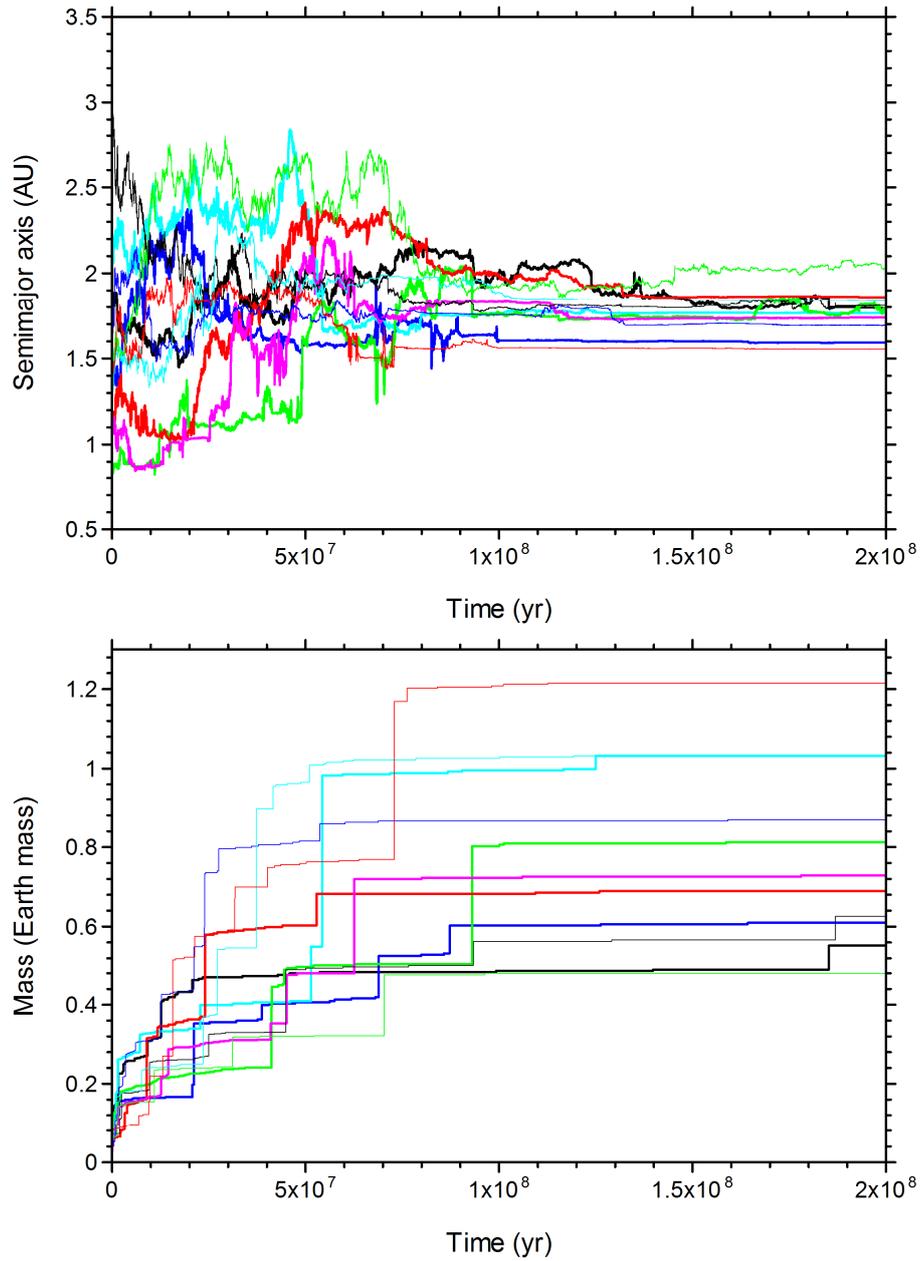

**Figure 6.** Orbital (top) and accretional (bottom) evolution of eleven Mars-like planets obtained in 11 simulation runs representing the non-migrating scenario, "Sim0" (Section 4.1). The planets are indicated with different colors or curve thicknesses when the color is the same for two objects.

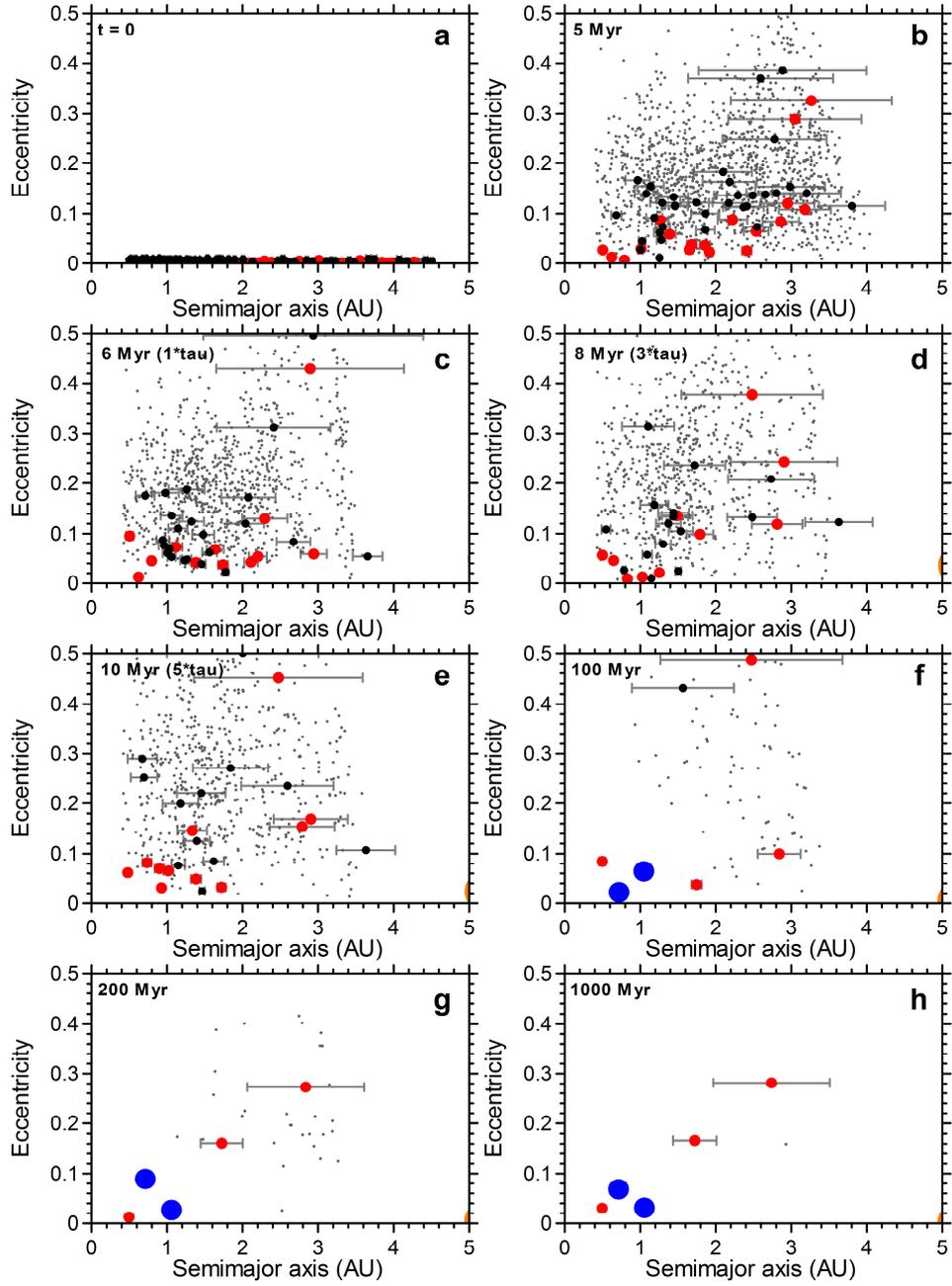

**Figure 7.** Time evolution of a typical migrating system, represented by run6 from simulation 4 (Table 1). Labels and symbols are the same as explained in caption of Fig. 3. See Section 4.2 for details.

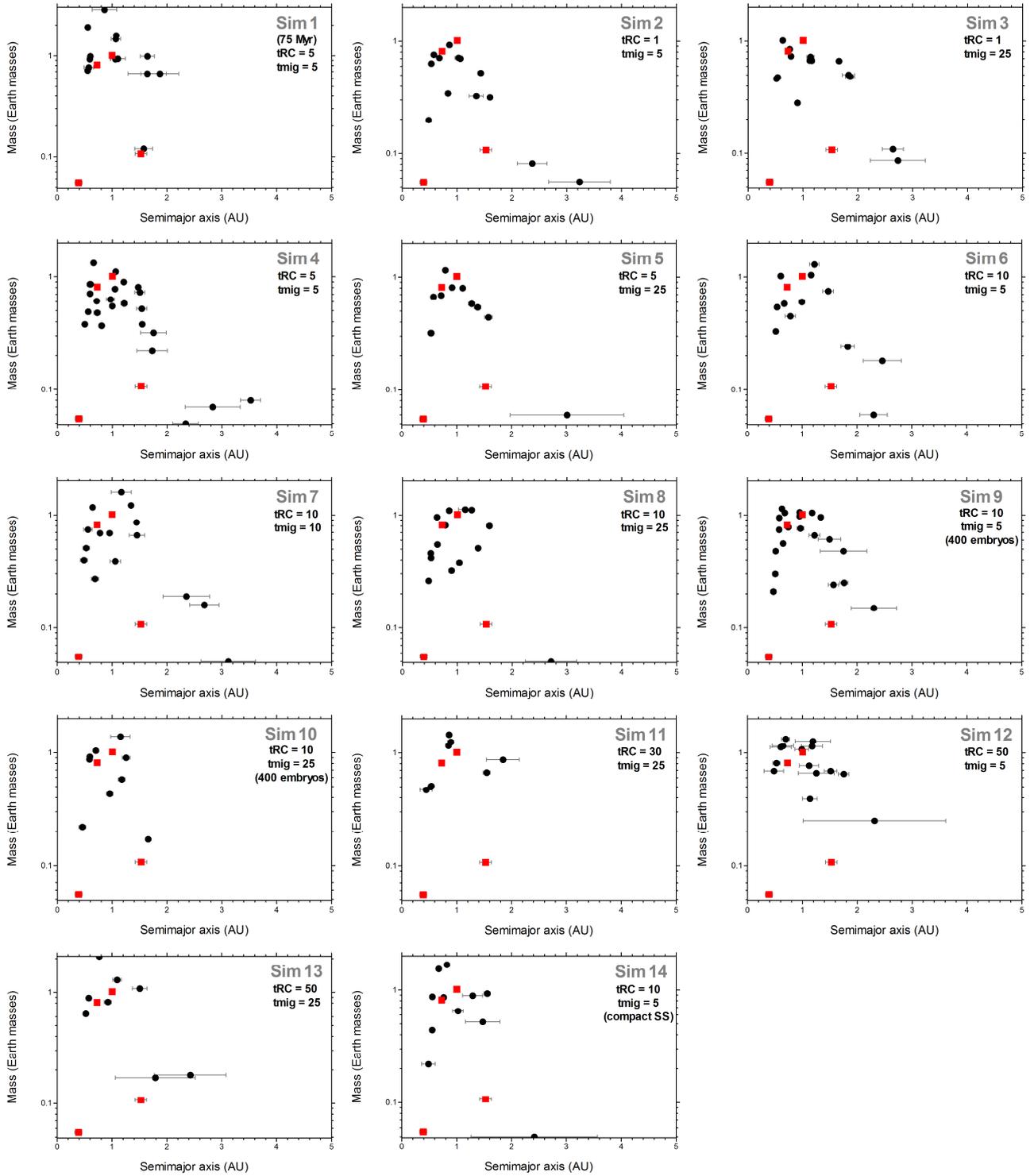

**Figure 8.** Planets obtained in simulations Sim1–14 of migrating systems after 200 Myr. Labels and symbols are the same as explained in the caption of Fig. 4. The timing of the 1:2 MMR crossing and the total migration timescale in Myr are given by $t_{RC}$ and $t_{mig}$, respectively. See Table 1 and Section 3 for more details. The results may be compared with those in Fig. 4.

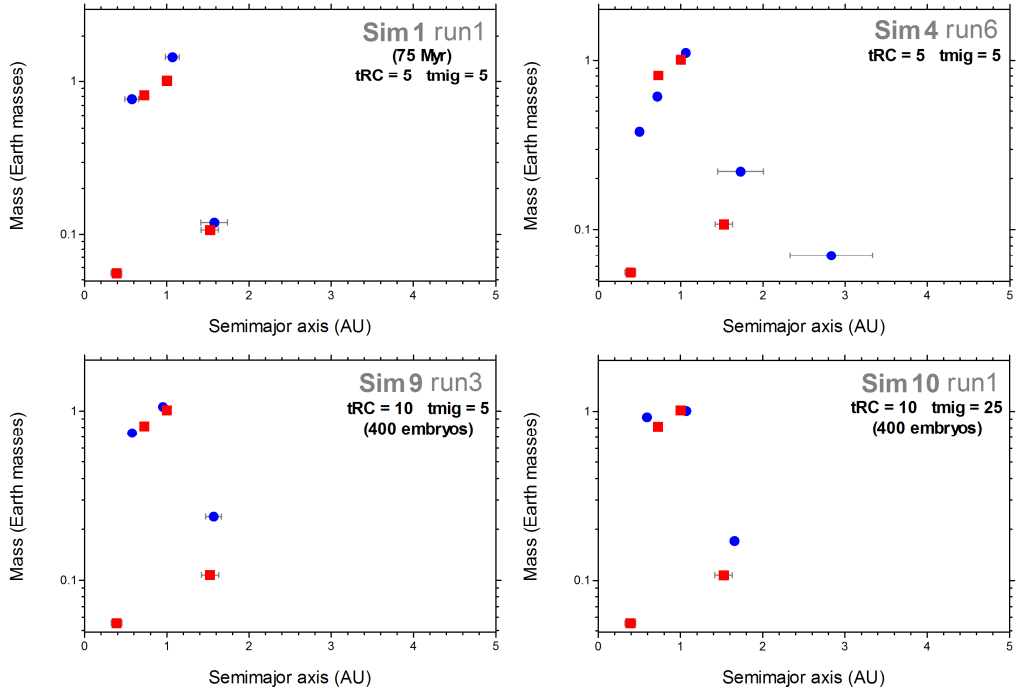

**Figure 9.** Terrestrial planets obtained in the best four individual runs of migrating systems after 1 Gyr, as indicated in the panel labels. The timing of the 1:2 MMR crossing and the total migration timescale in Myr are given by $t_{RC}$ and $t_{mig}$, respectively. See Table 1 and detailed discussion in Section 4.2.1 for details.

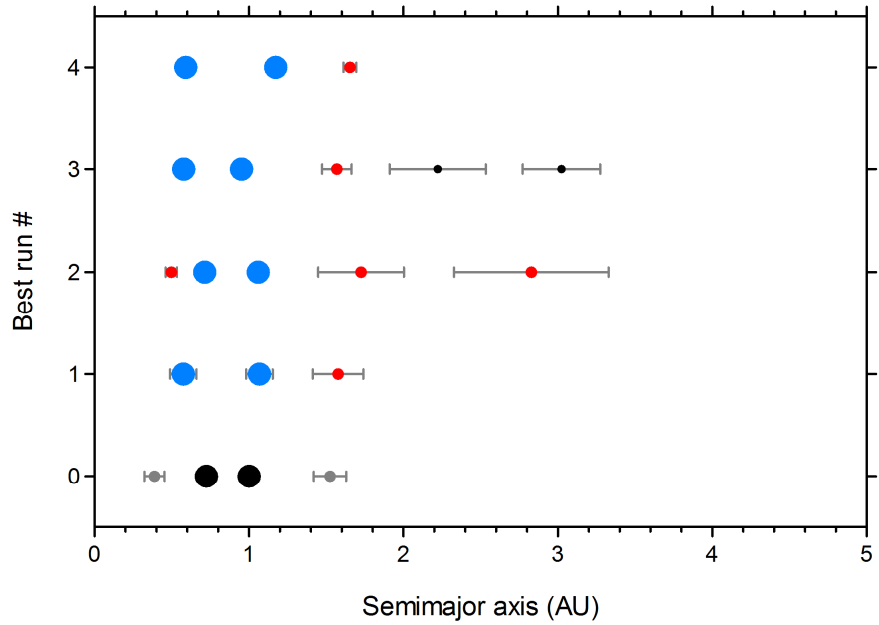

**Figure 10.** Obtained planetary systems after 1 Gyr for the best four individual runs of migrating systems. The best runs (in order 1-4) are represented by Sim1/run1, Sim4/run6, Sim9/run3, and Sim10/run1, respectively (as illustrated in Fig. 9). The labels and symbols are the same as in the caption of Fig. 5.

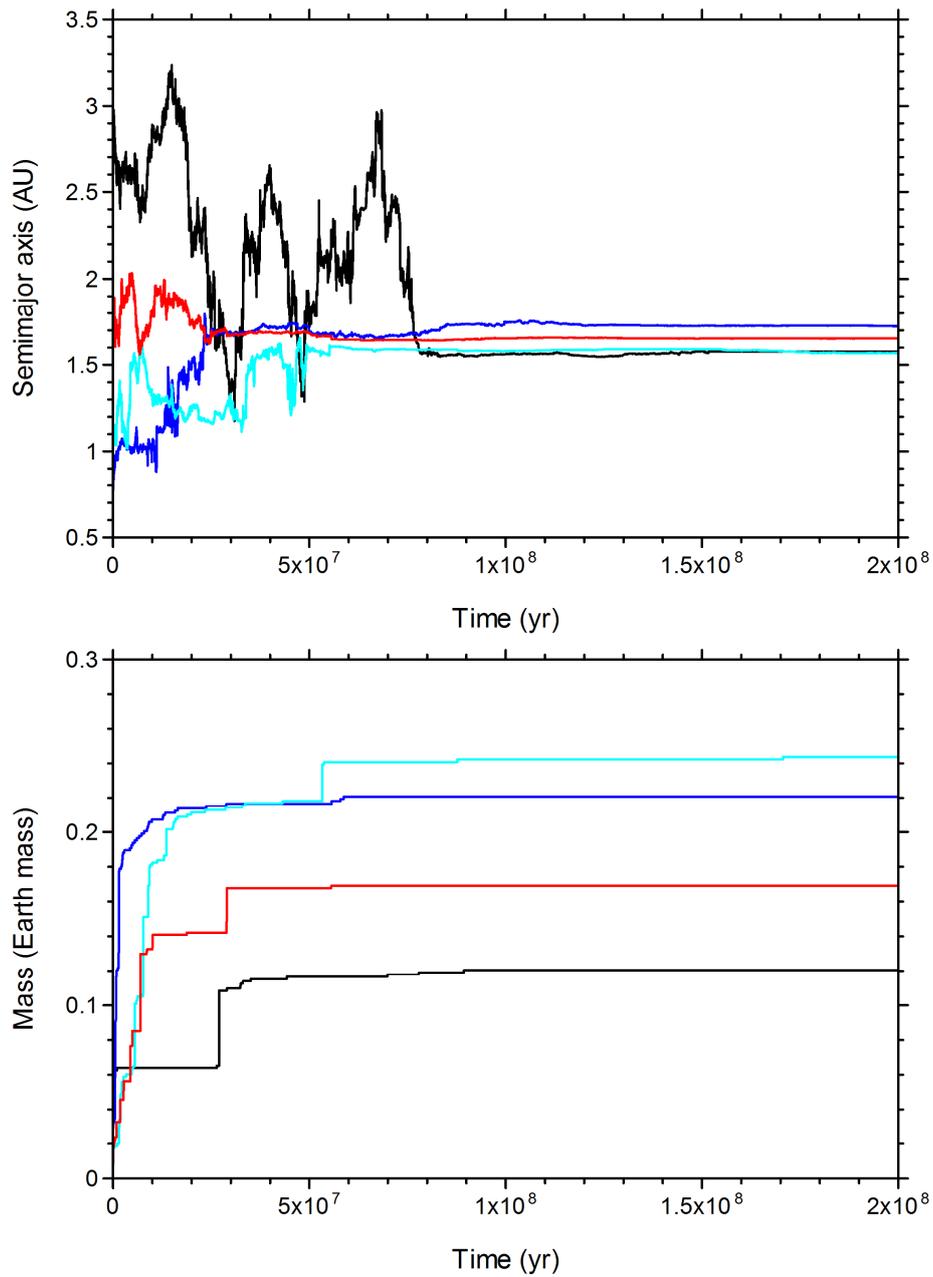

**Figure 11.** Orbital (top) and accretional (bottom) evolution of four Mars analog planets obtained in each of the best four individual runs of migrating systems, as illustrated in Fig. 10. The planets are indicated with different colors: black (best run 1), blue (best run 2), cyan (best run 3), and red (best run 4).

**Table 1.** Main simulations

| ID | Non-migrating $t$ (Myr)[a] | Pre-MMRc $t_{RC}$ (Myr)[b] | Post-MMRc $t_{mig}$ (Myr)[c] | Number of runs |
|---|---|---|---|---|
| 0 | 1000 | - | - | 12 |
| 1 | 75 | 5 | 5 | 6 |
| 2 | 0 | 1 | 5 | 3 |
| 3 | 0 | 1 | 25 | 3 |
| 4 | 0 | 5 | 5 | 6 |
| 5 | 0 | 5 | 25 | 3 |
| 6 | 0 | 10 | 5 | 3 |
| 7 | 0 | 10 | 10 | 4 |
| 8 | 0 | 10 | 25 | 4 |
| 9[d] | 0 | 10 | 5 | 6 |
| 10[d] | 0 | 10 | 25 | 3 |
| 11 | 0 | 30 | 25 | 3 |
| 12 | 0 | 50 | 5 | 5 |
| 13 | 0 | 50 | 25 | 3 |
| 14 | 0 | 10 | 5 | 4 |

We placed 100 embryos (2.7 $M_\oplus$ in total) and 2000 planetesimals (2.7 $M_\oplus$ in total) in a planetesimal disk at 0.5-4.5 AU. The system was compact, where $a_{J0}$ = 5.45 AU and $a_{S0}$ = 8.18 AU for simulations 1-13, and $a_{J0}$ = 5.4 AU and $a_{S0}$ = 7.3 AU for simulation 14. After Post-MMRc $t_{mig}$, all runs were followed until the total time span reached 1 Gyr.

[a] Timescale in which no planet migration was imposed on Jupiter and Saturn.
[b] Timescale until the MMR crossing event (modeled by a gentle linear migration until Jupiter and Saturn were on the verge of the crossing. See Section 3 for details).
[c] Migration timescale after the MMR crossing event, defined as $5\tau$, where $\tau$ is the exponential timescale described in Section 3.2.
[d] Used 400 embryos, instead of 100.